\documentclass{llncs}
\usepackage[utf8]{inputenc}

\usepackage{cite}
\usepackage{amsmath,amssymb,amsfonts}
\usepackage{algorithmic}
\usepackage{graphicx}
\usepackage{textcomp}
\usepackage{xcolor}
\def\BibTeX{{\rm B\kern-.05em{\sc i\kern-.025em b}\kern-.08em
    T\kern-.1667em\lower.7ex\hbox{E}\kern-.125emX}}
    
\usepackage{listings}
\usepackage{color}
 
\definecolor{codegreen}{rgb}{0,0.6,0}
\definecolor{codegray}{rgb}{0.5,0.5,0.5}
\definecolor{codepurple}{rgb}{0.58,0,0.82}
\definecolor{backcolour}{rgb}{0.99,0.99,0.98}
\definecolor{keyword}{rgb}{0,0,0.99}
 
\lstdefinestyle{Eiffelstyle}{
    backgroundcolor=\color{backcolour},   
    commentstyle=\color{codegreen},
    keywordstyle=\color{keyword},
    numberstyle=\tiny\color{codegray},
    stringstyle=\color{codegreen},
    basicstyle=\footnotesize,
    breakatwhitespace=false,         
    breaklines=true,                 
    captionpos=b,                    
    keepspaces=true,                 
    numbers=left,                    
    numbersep=5pt,                  
    showspaces=false,                
    showstringspaces=false,
    showtabs=false,                  
    tabsize=2
}   

\usepackage{multirow}

\usepackage{array}
\newcolumntype{L}[1]{>{\raggedright\let\newline\\\centering\arraybackslash\hspace{0pt}}m{#1}}

\title{Towards A Broader Acceptance Of Formal Verification Tools:}
\subtitle{The Role Of Education}

\author{Mansur Khazeev, Manuel Mazzara, Daniel De Carvalho, Hamna Aslam\\
\{m.khazeev, m.mazzara, d.carvalho, h.aslam\}@innopolis.ru}
\institute{Innopolis University, Russia}

\authorrunning{<abbreviated author list>}

\begin{document}

\maketitle

\begin{abstract}
Formal methods yet advantageous, face challenges towards wide acceptance and adoption in software development practices. The major reason being presumed complexity. The issue can be addressed by academia with a thoughtful plan of teaching and practise. The user study detailed in this paper is examining AutoProof tool with the motivation to identify complexities attributed to formal methods. Participants' (students of Masters program in Computer Science) performance and feedback on the experience with formal methods assisted us in extracting specific problem areas that effect tool usability. The study results infer, along with improvements in verification tool functionalities, teaching program must be modified to include pre-requisite courses to make formal methods easily adapted by students and promoting their usage in software development process.
\end{abstract}

\section{Introduction}  \label{Intro}

Professionals and academic researchers in the field of quality assurance have 
realized the benefits of static analysis and formal verification. Consequently, in the past decades, a significant effort has been put for developing and improving formal verification tools and related methodologies. While the extensive research conducted in recent years was mostly focusing on soundness and completeness of methodologies and tools (what can be verified and how), usability criteria were often disregarded. Aspects, such as user entry level, required background knowledge, and tools learnability was always considered secondary. It is understandable that a new technology requires a high push on the conceptual side to be delivered, however, without a better understanding of these aspects, it is unlikely that the verification tools could be widely adopted, especially in non-life-critical or mission-critical projects.  Usability should be considered as a significant quality attribute whenever the benefits of formal verification have to be exploited in practice.

This paper focuses on the the usability (defined in section \ref{Usability}) of a specific verification tool: AutoProof \cite{AutoProof2015}, a static analyzer for contracts defined for the Eiffel programming language. Eiffel allows the introduction of ``contracts" inside the code, following a software correctness methodology called Design By Contract (DbC) \cite{Meyer:1992b}. This approach uses preconditions and postconditions to document the change in state caused by a piece of a program. The usability of this particular tool has been assessed and discussed by Furia et al. \cite{Furia2015-AutoProofUsability}. However, in \cite{Furia2015-AutoProofUsability} the study by non-experts was based on group project, and more importantly the effort was not measured. The goal of this paper is to measure the effort that an average user can (and would be willing to) put to formally verify their program, and what the ``blockers" or ``hinderances" would be, in case, a user could not succeed. This is our endeavor and contribution on top of the existing literature.

There are two major reasons to consider AutoProof for this study. Firstly, Eiffel is an object-oriented programming language. The modularity and scalability that the object-oriented paradigm supports are highly valued in the developments of big software systems, thus widely used in modern software engineering\cite{pressman_software_2010}. Secondly, the Design By Contract (DbC)\cite{Meyer:1992b} methodology that was introduced almost thirty years ago, slowly but gradually has been adopted by many popular programming languages like Java (JML)\cite{jml_overview_2005}, C\# (Spec\#)\cite{specsharp}, C++ ($expects$ and $ensures$ clauses), Kotlin (preconditions) etc. We believe that the methodology implemented in AutoProof can be transferred to any other Object Oriented programming language without significant rework.

The paper is organized as follows: Section \ref{Usability} discusses the notion of usability and how it applies to our work. Section \ref{design} details study design, including comparison with similar studies, while Section  \ref{Results} shows the results of the study presented here. Section \ref{Issues} analyses the responses of participants and the outcomes. Students involved in the study were asked to write essays about the courses on formal methods; this is presented in Section \ref{attitude}. Finally, recommendations are developed in Section \ref{recommendations}.

\section{Usability} \label{Usability}
Usability is a broad concept that cannot be fully explored here. In this paper, we will focus specifically on some aspects of usability, and in particular applied to AutoProof. According to ISO standard\footnote{https://iso25000.com/index.php/en/iso-25000-standards/iso-25010/61-usability} usability is composed by several sub-characteristics of which we will analyze only the following:

\begin{itemize}
    \item \textbf{Learnability}: Degree to which a product or system can be used by specified users to achieve specified goals of learning to use the product or system with effectiveness, efficiency, freedom from risk and satisfaction in a specified context of use
    \item \textbf{Operability}: Degree to which a product or system has attributes that make it easy to operate and control
    \item\textbf{User Error Protection}: Degree to which a system protects users against making errors
\end{itemize}

Other aspects of usability such as appropriateness, recognizability and User interface aesthetics have been considered to have low relevance at the current stage, thus are not covered in this work.


In particular, in this work we will analyze the three following questions:

\begin{itemize}
    \item To what extent AutoProof supports Learnability?
    \item To what extent AutoProof supports Operability?
    \item To what extent AutoProof supports User Error Protection?
\end{itemize}



\section{Study Design} \label{design}


This section describe study design from the profile of participants to the tasks given, their duration and rationale. The survey was conducted as the warm-up exercise as a part of ``Requirements and specification" course at Innopolis University, Russia. 

\subsubsection*{Profile of Participants}

The study has been conducted with twenty-two participants of the Master Program in Software Engineering. All participants had background education in Information Technology from different universities all over Russia and some other countries such as Ecuador, Brazil etc. A distinguishing characteristic of this program is, enrollment require students to have at least three years of industrial experience. Therefore, we believe that knowledge level of the participants correspond to the level of junior developers from different companies not only in Russia. Nevertheless, non of them had an industrial experience on formal methods before joining the program and the only background knowledge they gained before participating in the survey was with Event-B \cite{Abrial:2010} and Promela \cite{Neumann2014} through one of the core courses of the program. 

\subsubsection*{Rationale for Choosing the Exercise}
During the introductory lecture for the current course, students demonstrated a poor understanding of the benefits formal methods might bring to software development process, in particular, during requirements specification phase. The instructor's motivation was focused towards learning outcomes rather than evaluation and therefore, exercise was deliberately chosen that had solution available publicly. This ensured less pressure on student's side for figuring out solutions and no grading anxiety. 
Students were explicitly informed that this individual assignment would not be graded and would be used only for ``warming-up" before a small group project. The students were instructed to review solutions only if they had no idea on how to proceed further in verification. They had to log all details, describing comprehensively, the issues they were trying to eliminate and the time needed to do so. Since, the students had no motivation to cheat, we believe, the data collected from this exercise is reliable. 

\subsubsection*{Assignment Description} The participants were given a home assignment to specify and verify any two of nine exercises from the tutorial available online \cite{AutoProof_tutorial} and, were asked to log their progress. They were also asked to write a short essay documenting their progress and their problems. The time for completing the assignment was not restricted. The students had to decide themselves on amount of effort they think is reasonable and they would spend on each exercise. The overall collection of data span over twenty-one days. The assignment was submitted using the Moodle platform \cite{dougiamas2003}.

\subsubsection*{Helping Material Provided} Before proceeding with the exercise the respondents attended a lecture where the essential concepts and principals for AutoProof tool were presented, as well as the proof process on particular examples was demonstrated and discussed. The students had free access to the documentation and all exercises together with the solutions.
\subsection*{Technical Background} \label{Technical usability}
There are a few technical concepts that are necessary to grasp in order to fully understand the exercises described below. For readers, with Software Engineering background, this part can probably be skipped.
In programming languages based on the notion of \emph{class} (\emph{object-oriented} programming languages), \emph{design-by-contract} is an approach that defines the ``rights'' and ``obligations'' of a class and its clients. During the execution of the program, one can enforce these rights and obligations to be checked: In case, one of these assertions is violated, an exception is raised, it is thus a way to find bugs in the program (and then to hopefully fix them!). Instead of checking  \emph{dynamically} (that is, \emph{during} the execution of the program), one can try to check \emph{statically} (that is, \emph{before} the execution of the program): One would like to \emph{prove} that, for \emph{any} execution of the program, none of the assertions specifying the rights and obligations might be violated. It is the purpose of the tool, AutoProof, to check statically these assertions, which are divided into the following types:
\begin{itemize}
\item \emph{Precondition} and \emph{postcondition} of a method/feature/function: The caller of a method/feature/function has the obligation to satisfy its preconditions, while the callee should guarantee its postconditions to hold at the end of its execution. 
\item \emph{Class invariant}: Class invariants define properties of the state of an object that should be preserved during the execution of the program.
\item \emph{Loop invariant}: It is an assertion that should be maintained at each iteration of the loop.
\item \emph{Framing assertion}: Framing assertions express which objects can have their state modified.
\end{itemize}
Furthermore, a \emph{loop variant} is a measure that should decrease strictly at each iteration of the loop to ensure that the loop eventually terminates.

\subsection*{Exercises Description} \label{Ex_discr}
The tutorial used for the survey was prepared by the developers of the AutoProof with a purpose of giving an overview of the tool and demonstration of how to verify Eiffel programs. It includes generic and common examples such as BANK\_ACCOUNT, CLOCK, several searching and sorting algorithms and data structures - 11 in total. The students were allowed to pick any two examples except “Account” (it was covered in class) and “Max in array” (which is already complete). All of them capture different verification challenges thus differ in complexity, however all could be mapped into three categories of increasing complexity: verification of basic properties, verification of algorithms, and object consistency and ownership. The students have tackled the exercises only from first and second category. The comparison of the exercises that were attempted by students to verify is presented in Table \ref{exer-comp}, where ``Implementation" is the implementation size in LOC, ``Specification" is the number of expected annotations for specification (contracts) and ``Verification" denotes a number of verification specific annotations necessary to complete the exercise.
\begin{table}
\centering
\begin{tabular}{|L{1.4cm} |L{2.4cm} |L{2cm} | L{1.8cm} | L{4cm} |}
\hline
Exercise & Implementation & Specification & Verification & Required knowledge \\ \hline
Clock & 56 & 24 & 11 & model query, wrapping, framing \\ \hline
Linear search & 14 & 6 & 0 & model query \\ \hline
Binary search & 24 & 13 & 0 & model query \\ \hline
Gnome  sort & 25 & 25 & 7 & model query, wrapping, framing, functional feature, termination \\ \hline
Insertion  sort & 33 & 37 & 9 & model query, wrapping, framing, functional feature \\ \hline
\end{tabular}
\caption{Comparison of the Exercises}
\label{exer-comp}
\end{table}
\subsubsection{Basic Properties: the ``Clock" exercise}
One of the exercises that was chosen by most of the student was ``clock". The class CLOCK in this example, implements a clock with $hours$, $minutes$ and $seconds$; has features to set values to these attributes and increase their current value by one: $increase\_hours$, $increase\_minutes$ and $increase\_seconds$, and a constructor feature to instantiate an object of this type. The implementation was already given, the task was to specify invariants of the class and pre-/post-conditions for each feature so that AutoProof can verify correctness of an implementation provided. 
\lstset{style=Eiffelstyle}
\renewcommand\thelstlisting{\arabic{lstlisting}}
\setcounter{lstlisting}{0}
\begin{lstlisting}[caption={Partial Solution for ``clock" Exercise},label={clocksolution},language=Eiffel]
class CLOCK
. . . -- Some features were omitted
feature
increase_minutes
    -- Increase `minutes' by one.
  note
    explicit: wrapping
  require
    modify_model (["minutes", "hours"], Current)
  do
    if minutes = 59 then
      set_minutes (0)
      increase_hours
    else
      set_minutes (minutes + 1)
    end
  ensure
    minutes_increased: 
    minutes = (old minutes + 1) \\ 60
    hours_unchanged: 
    old minutes < 59 implies hours = old hours
    hours_increased: 
    old minutes = 59 implies hours = (old hours + 1) \\ 24
  end

invariant
  hours_valid: 0 <= hours and hours <= 23
  minutes_valid: 0 <= minutes and minutes <= 59
  seconds_valid: 0 <= seconds and seconds <= 59
end
\end{lstlisting}
A code snipped with partial solution for ``clock" exercise is depicted in the Listing \ref{clocksolution}. It includes only one feature and class invariant.
A feature $increase\_minutes$ is equipped with the precondition, which includes only a frame condition, denoting what attributes are permitted to be modified (line 6) and the post-conditions describing how these attributes should be modified (lines 15-17). A keyword $old$ is used to refer to the value of a variable before execution of the feature body. Post-conditions are: (line 15) minute should increase by one but rounded by 60; $hours$ should stay the same (lines 16) or increase by one (lines 17) depending on $old$ value of attribute $minute$. Class invariants (lines 21-23) describe valid values for $hours$, $minutes$ and $seconds$.

\subsubsection{Searching and Sorting Algorithms}

Other exercises, chosen by several students were verification of some basic searching (linear, binary) and sorting (insertion, gnome) algorithms. In these exercises, student were verifying, not classes but functions containing an iteration or a recursive call. Similar to ``clock" exercise, implementation of the algorithms was provided. The challenge was to find appropriate invariants and variants of loops, in addition to pre-/post-conditions of the functions. In case of searching algorithms, there was no additional annotation needed for succeeding in verification, however, it was necessary to understand the notion of model queries. In verification of sorting algorithms, the students had to take care of frame conditions, control the consistency of the objects by means of unwrapping and wrapping as well as to understand difference between specification specific features (depicted in Table \ref{exer-comp}).

When everything done correctly, the students had to get all the features of the class verified. In addition, all exercises contained a client (another class) to check the correctness of specification of initial class.

\section{Verification Process and Collected Data} \label{Results}


The students were asked to record their verification process in a free format. A key aspect of this exercise was to write down the encountered difficulties and track the time that was spent to overcome them. Many participants have created a very detailed log, in some cases, with screen shots or code snippets. Several reports were brief and less informative, however, could capture the main blockers and total time spent. Only one participant did not log the time, they spent on the exercise. Nevertheless, they had mentioned the main issues that were encountered. We have, therefore, considered this data as well in the analysis. 

The data extracted from the reports submitted by the students showed that about 2/3 of the group started from the very basic exercise ``Clock", and as second exercise, they tried to verify ``linear search algorithm".
\subsection{Student Performance Results for ``Clock" Exercise} 
The ``clock" exercise was attempted by seventeen students. Table \ref{clock-results} shows the analysis and the comparison, regarding the time spent by the students, trying to specify and verify this exercise. 
\begin{table}
\centering
\begin{tabular}{|L{2cm} |L{1.5cm} |L{1.5cm} | L{1.5cm} |}
\hline
    & Succeeded & Failed  & Total      \\ \hline
Min     & 19    & 69    & 19  \\ \hline
Max     & 630   & 180   & 630 \\ \hline
Mean    & 180   & 150   & 170 \\ \hline
\end{tabular}
\caption{Time Spent on ``clock" Exercise (in minutes)}
\label{clock-results}
\end{table}

The average time spend by the students completing the verification of the class CLOCK successfully (180 minutes) differ from time spend by those who failed (150 minutes) by 30 minutes. However, there is an enormous difference between maximum and minimum values: the quickest student finished the exercise in 19 minutes, however, for the most persistent, it took more than 10 hours to finish, although it was done with breaks. The students, who could not complete this exercise usually gave up in no more than three hours.

\subsection{Cumulative Results}

Table \ref{numeric-results} summarizes the results for all the exercises of the ``warm-up session", separately depicting the number of students succeeding as well as failing each exercise. Additionally, the table also gives the total number of students tackling each exercise.

\begin{table}
\centering
\begin{tabular}{|L{2cm} |L{1.5cm} |L{1.5cm} | L{1.5cm} |}
\hline
            & Succeeded & Failed  & Total \\ \hline
Clock          & 11        & 6    & 17    \\ \hline
Linear search  & 10        & 9    & 19    \\ \hline
Binary search  & 2         & 3    & 5     \\ \hline
Gnome sort     & 1         & 1    & 2     \\ \hline
Insertion sort & 0         & 1    & 1     \\ \hline
\end{tabular}
\caption{Number of Students per Exercise}
\label{numeric-results}
\end{table}

The number of students who succeeded in the ``clock" exercise (11 students) almost twice as many as those who could not verify it (6 students). However, in the exercises with an iteration through a collection to find an element the numbers became roughly equal: 10 students succeeding and 9 failing verification of linear search algorithm, 2 succeeding and 3 failing verification of binary search algorithm. Only one out of three students tackling verification of sorting algorithms managed to complete it successfully.
\section{Study Results and Issues Identified} \label{Issues}

The most significant results from the acquired data are blockers, during the verification process, i.e. problems that the students could not overcome - blockers. The identified blockers are separated into two categories, those related to \emph{specification} and those to \emph{verification}. The summary is shown in Table \ref{blocker}.

\begin{table}
\centering
\begin{tabular}{|l|l|c|}
\hline
Category & Subcategory & Occurrence \\ \hline
\multirow{2}{*}{Specification} & Precondition & 2 \\ \cline{2-3} 
 & Postcondition & 7 \\ \hline
\multirow{4}{*}{Verification} & Loop invariant & 9 \\ \cline{2-3} 
 & Framing & 7 \\ \cline{2-3} 
 & Model query & 7 \\ \cline{2-3} 
 & Wrapping & 1 \\ \hline
\end{tabular}
\caption{Blockers: Tasks in the Exercises, the Students could not Resolve}
\label{blocker}
\end{table}

\textit{Specification blockers} are any issues related to expressing conceptual properties in the contracts: typically pre- and post-conditions. In general, class invariants should also be included in this list. However, all participants attempting to verify the ``Clock" exercise were able to express its invariant. Other exercises did not require to specify class invariants. The students often had problems with expressing preconditions such as ``array is sorted" for the binary search algorithm, or failed to prove the client classes, as they had specified weak post-conditions for feature in the initial class. This is defined as ``silence" \cite{formalism_1985} in specifications. A typical example from students' responses is missing to denote behaviour of $hours$ in the post-condition feature of $increase\_seconds$.

Other blockers are \textit{verification}-related. Category comprises of issues to express loop invariants (and variants), specification of frame conditions, use of model queries and ``play" with objects consistency and ownership (although such exercises were not considered by the students). This category describes all blockers that are specific to the methodology implemented in the tool.

The categories presented in Table \ref{blocker} are in relation with the subcategories of usability described in Section \ref{Usability}:

\subsubsection{Learnability issues leading to Specification Problems} include difficulty of grasping fundamental concepts, needed for using the tool such as predicate and hoar logic etc. These issues in a sense are the prerequisites for being able to use the tools, and there is no way of resolving them apart from conducting a training.
\subsubsection{Learnability issues leading to Verification Problems} include difficulty of grasping techniques used in the tool to resolve certain verification challenges. For example, the tool requires manual identification of frame conditions for every feature, which turned out to be a blocker for 7 exercises. Some of the issues that are methodology dependent, and could be eliminated by improving the methodology implemented in the tool. Others, are still related to fundamental concepts, such as understanding proof by induction to be able to provide strong loop invariant to prove an iteration.
\subsubsection{Operability issues leading to Specification Problems} include minor issues of the tool such as not supported all the language constructs. For example, use of \textbf{across} to iterate some data structure is not supported, and can be rewritten using other constructs. Tool notifies the user about it, however still consumes some effort and time of users to resolve it. These are mostly intentional technical debt \cite{technical_dept} that the developers have to pay to improve the usability of the tool. 
\subsubsection{Operability issues leading to Verification Problems} are related to a difficulty of understanding the messages returned by the tool in case of verification failure. Often the feedback of the tool is not much helpful to clarify whether the specification is incomplete or implementation does not confirm to it. This comes from the fact that the methodologies implemented in the tool are necessarily incomplete, since they target undecidable problems \cite{polikarpova_specified_2014}. Nevertheless, these type of issues could be partially eliminated by improving methodologies and using additional techniques, for example, utilizing symbolic execution\cite{cadar_symbolic_2013} for generating counter examples.
\subsubsection{Error Protection issues leading to Specification Problems} include inability of the tool to detect any contradictions in the specifications. For example, contradiction in the precondition of the feature may lead to its successful verification, whereas there can not be any use of such feature. 
\subsubsection{Error Protection issues leading to Verification Problems} comes from the complicated methodology of the tools that includes many build-in attributes misusing of which might result in wrong verification success. Current methodology definitely should be improved in a way that it should not allow user to hack the verification process.

\section{Attitude Towards the Adoption of Formal Methods}
\label{attitude}
The participants of the study were asked to write an essay about the courses in formal methods they have attended. Most of them considered hard the use of formal methods. We are summarizing here the reasons that have been identified.

\begin{itemize}
\item \textbf{Learnability issues}: It is apparent that students were not sufficiently strong in math and logic. For instance, one participant reported:\textit{ ``And I really do not understand why tools consider differently some conditions like: Result = 0 implies not a.sequence.has(value) and not a.sequence.has(value) implies Result = 0. For me both of them are the same"}. This feedback seems to be the consequence of a fundamental lack of prerequisites in (propositional) logic.

\item \textbf{Operability issues}: One participant reported  \textit{``Of all the tools we have used only SPIN is convenient and pleasant to work with; others lack documentation and are not stable enough.''}. This is clearly a problem of the tools, not of the users.

\item \textbf{Error Protection issues}: One participant reported, for instance: \textit{``How can I be sure that the prover has no errors in it and really verifies my code in a good manner?''}. This is a classic meta-problem and it applies to other software artifacts such as compilers and operating systems. Specifically, AutoProof does not seem to be mature enough for these kind of users. The tool may at times report that the code has been verified while, in fact, the verification of some assertions have been disabled.
\end{itemize}

Furthermore, several students were not really motivated and believed that verification methods are time consuming and not effective. One participant reported : \textit{``I did not understand where I can apply this approach in practice."} This demands for changes in the way formal methods courses are delivered. One possibility is, instead of presenting approach and tool at the same time with the problems they are supposed to solve (``find whether there are bugs and where they are"), they should be presented \emph{after} asking students to find the bugs in the program. In this case, students could compare two situations: the one in which they had to find bugs \emph{without} the tool (only reading the code and testing the program), and the one where they could use the tool. This approach would help realizing that trying to fix bugs without any support is less efficient and more time consuming than using the tool.

\section{Recommendations to Facilitate Formal Methods Adaptation through Academia} \label{recommendations}
Students feedback and the analysis of their performance suggest that formal methods require more attention from the education sector in order to be widely adopted in practice. Verification tools must be presented and discussed by offering an understanding of the implementation techniques, and guidance upon the verification process. Teaching DbC is necessary \cite{teaching_dbc}. Furthermore,\emph{Design by Contract} and \emph{Predicate Logic} are crucial prerequisites for getting students ready to dive into verification.

The current limitation of the tools are compelling issues in general, and in particular, if we refer to AutoProof, which is the target of this study, we can answer the questions highlighted in Section \ref{Usability} through the analysis of current user study. 
\begin{enumerate}

\item To what extent AutoProof supports Learnability? \newline
    -Although AutoProof uses code contracts written in the same programming language, the tool require training to get user familiarized with the techniques incorporated in the tool: it requires clear knowledge of the sophisticated methodologies implemented under the hood. However, once the user is acquainted with the tool environment and got experience, the usage becomes clear. However, these methodologies could be adopted only by advanced user and experts in the field.

\item To what extent AutoProof supports Operability?\newline
    -The concern in terms of operability is the lack of clarity of error messages as mentioned in Section \ref{Issues} and necessity in a number of additional annotations. This is the major problem area as identified to be causing complications towards tool operability.
    
\item To what extent AutoProof supports User Error Protection?\newline
    The compelling issue identified is to block possibilities of manipulating the verification process, which at presently tool lacks.
    \end{enumerate}
    


Summarizing:

\begin{itemize}
    \item Even simple exercises require significant effort (surely having more experience, the students might perform faster, however, the complexity of problems could be different as well)
    \item In its current state, the tool can not be used by ordinary programmer, usability should be reconsidered
    \item The educational delivery needs to be improved in order to provide students with better motivation: ``let us struggle without the tools first, and then struggle with the tool"
\end{itemize}
The present conclusions are drawn from the user study including twenty-two participants. We believe that further studies with different settings such as, examining users who have a strong background in predicate and logic concepts (especially taught as prerequisites for formal methods), will give useful insight into tool's strength and deficiencies. 
\bibliographystyle{splncs04}
\bibliography{bibliography}
\end{document}